\documentclass[english]{article}
\usepackage[T1]{fontenc}
\usepackage[latin9]{inputenc}
\usepackage{geometry}
\usepackage{color}
\usepackage{float}
\usepackage{graphicx}
\usepackage{amsmath}
\usepackage{amssymb}
\usepackage{babel}


\usepackage{babel}
\begin{document}

\title{An algorithm for minimization of quantum cost}

\author{Anindita Banerjee and Anirban Pathak}

\maketitle
\begin{center}
Jaypee Institute of Information Technology University, Noida, India
\par\end{center}


\begin{abstract}
\noindent A new algorithm for minimization of quantum cost of classical reversible and quantum circuits
have been designed.  The quantum costs obtained using
the proposed algorithm is compared with the existing results and it
is found that the algorithm  produces minimum quantum cost in all cases.

\medskip

\noindent {\bf Keywords:} quantum cost, circuit optimization and quantum circuit.
\end{abstract}

\section{Introduction}

According to Landauer's principle \cite{Landuer} any logically irreversible operation on information, is always
associated with a loss of energy. For example, each bit of lost information leads to the release of at least
kTln2 amount of heat. This type of energy loss is expected to become a substantial part of energy dissipation in
VLSI circuits in near future. The energy dissipation problem of VLSI circuits can be circumvented by using
reversible logic because reversible computation does not require to erase any bit of information. This
observation has  motivated scientists to design  reversible  circuits for various purposes \cite{Kerntopf}-\cite{Takahashi}. If the reversible circuit implements  quantum computation and it
comprises of quantum gates then it is  a quantum circuit and if  the  reversible circuit  implements only
classical computation  (boolean logic) then it is a classical reversible circuit. In the area of quantum
computing several new possibilities appeared which are impossible in classical domain. To be precise, quantum
teleportation \cite{teleportation}, infinitesimally secured cryptography \cite{cryptography1} and super
dense-coding \cite{superdensecoding} do not have any classical analogue. All these unique features of quantum
communication are associated with some circuits which are reversible in nature. In other words, we require
quantum circuits to implement quantum algorithms and protocols. For example, circuits are proposed for
implementation of  Shor's algorithm \cite{shor-QEC}, quantum teleportation \cite{teleportation}, various attacks
on quantum key distribution protocols \cite{cryptography1, cryptography2}, super dense coding
\cite{superdensecoding}, quantum error correction \cite{Steane-QEC, calderbank-QEC}, fault tolerant quantum
computation \cite{Shor-Fault}-\cite{ Gottesman-fault}, Grover's algorithm \cite{Grover}-\cite{implementinggrover2}, nondestructive discrimination of Bell states \cite{M.-Gupta}, quantum
circuits for addition  \cite{Takahashi} etc. Here we would like to note that  all quantum mechanical operations
are reversible and the only  difference between a classical reversible gate and a quantum gate is that the
classical reversible gate can not handle superposition of states (qubit). Consequently, set of all classical
reversible gates form a subset of set of all quantum gates. For example, Cnot gate can be achieved in classical
and quantum domains but the Hadamard gate can be achieved in quantum domain only. Therefore, classical
reversible circuits are only a subset of quantum circuits and any protocol designed for optimization of
particular parameter related to quantum circuits will also be valid for classical reversible circuits.

We have tried to explain the requirement and beauty of quantum circuits and now  the question arises: How to
obtain these circuits? There exist several algorithms for synthesis of classical reversible circuits
\cite{Kerntopf, Agrawal, Miller, Gupta, Ahmed} and quantum circuits \cite{Shende}-\cite{Ruican}. But these
algorithms do not provide a unique output. For example, a reversible multiplier can be achieved in many ways
\cite{multiplier1}-\cite{Banerjee}. Therefore, a
quantitative measure of the quality of a circuit is required. Some of the important quantitative measures are
gate count, number of garbage bits and quantum cost. Gate count is the total number of gates in a circuit, but
there is a specific problem with this quantitative measure of  circuit quality. Specially it is not unique. If
one is allowed to introduce a new gate or a complex gate library then the gate count can be considerably
reduced. An n-qubit reversible gate is represented by $2^n\times2^n$ unitary matrix. Product of any arbitrary
number of unitary matrices is always unitary. Moreover serial connection of such gates correspond to
multiplication of their matrices and parallel connection corresponds to tensor multiplication of their matrices.
Therefore, if we put a set of reversible quantum gate in a black box then it can be visualized as a new gate.
Thus the gate count can be reduced to 1. For example in \cite{Maslovbenchmark} the circuit cost of a
full adder circuit from NCT%
\footnote{NCT gate library is a universal gate library \cite{Maslovbenchmark}
comprising of NOT, Cnot and Toffoli gates.%
} gate library is 4, in \cite{multiplier5} it is reduced to 2 by using Peres gate and in
\cite{multiplier4-Islam} it is reduced to 1 by using a new gate. All the differences in circuit cost of full
adder is because of choice of non unique gate libraries. Consequently it is important to define an unique gate
library for comparison of circuit cost. Further, a good quantum circuit requires minimum number of garbage bits.
This is so because garbage bit is defined as an additional output bit which is required to make a function
reversible and it is not used for further computations.  The quantum cost%
\footnote{Definition of quantum cost is discussed in detail in section 2.%
} \cite{Mohammadifiguesofmerit, Smolin, Barenco} of a reversible circuit is the number of primitive quantum
gates needed to implement a circuit. Primitive quantum gates are the elementary building blocks \cite{Barenco}-\cite{Maslov2}, like Not gate, Cnot gate, controlled-$v$, controlled-$v^+$, rotation gates etc. We can
construct Toffoli gate with square root of Not gate (V)  and Cnot gate and in that construction the minimum gate
count of Toffoli gate is 5 \cite{Smolin} and its  quantum cost is also 5. These requirements yield separate
measures of quality of a quantum circuit. To be precise, the circuit is  better if it has lesser number  of
garbage bits, circuit cost and  quantum cost. But it is often observed that reduction of circuit cost leads to
increase in garbage bits and reduction of quantum cost leads to increase in circuit cost \cite{multiplier5}.
Keeping these in mind we have recently introduced a new parameter called Total cost (TC) \cite{Banerjee} which
is the sum of gate count of an optimized circuit, number of garbage bits and quantum cost. For reduction of TC
it is required to simultaneously reduce the circuit cost, garbage count and quantum cost. This is an open
problem and at present neither an algorithm for simultaneous reduction of all  these measures nor a satisfactory
algorithm for reduction of quantum cost exist. Before we address the more complex problem of minimization of TC
we have to device a protocol for reduction of quantum cost. In some works \cite{Mohammadifiguesofmerit, Gupta,
multiplier4-Islam, Revlib} the quantum cost is calculated straight by adding the quantum cost of respective
reversible gates in the circuit or it is optimized by applying deletion rule only. A simple minded systematic
approach is also proposed by Maslov \emph{et al.} \cite{Maslov2, Maslov1}. These facts have motivated us to
design an algorithm for minimization of quantum cost and to apply our algorithm to compute quantum cost of
different circuits in \cite{Mohammadifiguesofmerit, Gupta, Maslovbenchmark, Maslov2, Revlib}.

In the next section we have discussed the earlier approaches and their limitations. In section 3, we have
proposed an algorithm for calculating the quantum cost of  reversible  circuits. Here we have  also compared our results with the earlier proposals \cite{Mohammadifiguesofmerit, Gupta, Maslovbenchmark, Maslov2, Revlib} to establish that
the quantum cost computed by the proposed algorithm is minimum. Finally we conclude in section 4.
\section{Previous works}

Cost of an arbitrary unitary gate was first introduced by Barenco \emph{et al.} in 1995 \cite{Barenco}. They had
considered all $2\times2$ gate and Cnot gate as basic gates and had shown that for
any $2\times2$ unitary gate U%
\footnote{n qubit gate is represented by a $2^{n}\times2^{n}$ unitary matrix. Therefore a $2\times2$ and
$4\times4$ gates correspond to 1 qubit and 2 qubit
gates respectively. Different notations have been used in \cite{Mohammadifiguesofmerit, Miller, Perkowski, Hung, MohammadiHeuristic}. %
}, we can realize C-U (corresponding controlled U gate) by using at most 6 basic gates. But to analyze the cost
of a large gate (n bit Toffoli) he had considered the cost of C-U as $\Theta(1)$. Next year, Smolin and
DiVincenzo \cite{Smolin} calculated cost of Fredkin gate. In their calculation they went beyond the definition
of Barenco \emph{et al.} and assumed that the cost of every $4\times4$ gate is 1. This consideration does not
have any contradiction with Barenco \emph{et al}.'s definition of cost, as cost of all 2 qubit quantum gates is
$\Theta(1)$. Further progress in cost calculation was made by Perkowski \emph{et al.} \cite{Perkowski} in 2003
where they show that a one qubit gate costs nothing, if it precedes or follows by a 2 qubit gate. This is so
because one qubit gate can be combined with the 2 qubit gate to yield a new two qubit gate. Thus, the cost is
calculated as a total sum of $4\times4$ gates used. Following this definition  the cost of swap gate is one and
that of Peres gate is four. Peres gate is universal for reversible boolean operations and it has the minimum
cost compared to other universal gates. This observation of Perkowski \emph{et al.}  had motivated others to use
Peres gate to minimize the cost. For example, Maslov and Duek \cite{Maslov-improvedQC} have used the idea of
Perkowski \emph{et al.} and have shown that, the number of elementary quantum operations required to implement
Peres gate is less so it can be substituted for n-bit Toffoli network to reduce the cost of n-bit Toffoli gates.
Here we would like to note that in the earlier works \cite{Smolin,  Barenco, Perkowski} quantum cost was
mentioned as cost. The term quantum cost was coined by Maslov \emph{et al}. \cite{Maslov-improvedQC,
Maslov-thesis} in 2003, they have defined quantum cost of a gate G, as the number of elementary quantum
operations required to realize the function given by G. Later on, Hung \emph{et al.} \cite{Hung} had
reconsidered the quantum cost estimation protocol defined by Smolin and DiVincenzo \cite{Smolin}. They have
stated that each two qubit gate and each symmetric gate pattern (see Fig. 2 of \cite{Hung}) have quantum
implementation cost 1. In essence all these definitions of quantum cost are synonymous and we can follow
Perkowski 's definition \cite{Perkowski} and state that the quantum cost of a classical reversible or quantum
circuit is the minimum number of  one qubit and two qubit quantum gates needed to implement the circuit.

In recent past quantum cost of different classical reversible and quantum circuits have been reported \cite{Agrawal, Mohammadifiguesofmerit,
Gupta, multiplier4-Islam, Maslovbenchmark, Maslov2, Revlib, MohammadiHeuristic}. Simultaneously several efforts
have been made to reduce the quantum cost of different gates/circuits. For example, Barenco \emph{et al.}
\cite{Barenco} estimated the cost of a 6 bit Toffoli gate as 61.  Maslov and Duek \cite{Maslov-improvedQC}
reduced the quantum cost of this gate initially to 48 by using Peres gate. Further Maslov \emph{et al.}  \cite{Maslov1} reduced the quantum cost of this gate to 38 by
applying local optimization tools. There also exist  following two online databases: i) benchmark page of Maslov \emph{et al.}
\cite{Maslovbenchmark} and ii) Revlib \cite{Revlib},  which include quantum cost of different
circuits calculated by different authors.
In 2005 Maslov \emph{et al.} \cite{Maslov1} have shown that a closer look into
the cost metric can classify them into two subclasses: linear cost (where the quantum cost of a circuit is
calculated as sum of quantum cost of each gate) and nonlinear cost (where local optimization algorithm is used). According to this classification
scheme \cite{Maslov1} quantum cost defined in Smolin and DiVincenzo \cite{Smolin} and Hung \emph{et al}.
\cite{Hung} is nonlinear.
Interestingly, Haghparast and Eshghi \cite{Mohammadifiguesofmerit} have given following two prescriptions for
calculation of quantum cost:
\begin{enumerate}
\item Implement a circuit/gate using only the quantum primitive $\left(2\times2\right)$
and $\left(4\times4\right)$ gates and count them
\item Synthesize the new circuit/gate using the well known gates whose quantum
cost is specified and add up their quantum cost to calculate total
quantum cost.
\end{enumerate}

In both of these cases we will obtain linear cost metric and consequently the quantum cost obtained in these two
procedures may be higher than the actual one unless local optimization algorithms are applied to the entire
circuit. When we apply the local optimization algorithm on the entire circuit then we obtain a nonlinear cost
metric. The proposed algorithm will calculate nonlinear cost metric.  Local optimization is expected to  play an important role in minimization of quantum cost. Maslov \emph{et al.} \cite{Maslov2, Maslov3, Maslov4} have realized this fact
and have proposed an algorithm for minimization of quantum cost by applying templates and it yields nonlinear cost metric.

Our current work and work of Maslov \emph{et al.}'s \cite{Maslov2} is contemporaneous and independent. They
differ greatly in their premises, methods and consequences. 1) Maslov \emph{et al.}'s work deal with circuit
optimization precisely minimizing gate count by local optimization tools. They have introduced  templates and
applied them to optimize  the gate count. In contrast, our algorithm exploits a conceptual difference between
optimization algorithm used for reduction of gate count and the one used for reduction of quantum cost. 2) They
are restricted to a particular gate library  but to reduce the quantum cost we have introduced  new gates as
long as the gate is $2\times2$ or $4\times4$ quantum gate. 3) In their work the local optimization tools reduces
the gate count only, but in our work it is applied to reduce the quantum cost as well. This is shown in Fig. 2c
where moving rule \cite{Maslov2} (which was essentially designed to reduce circuit complexity) has not reduced
the circuit complexity but has reduced the quantum cost. This is evident in the work of Smolin \cite{Smolin}.
Further, we would like to note that the quantum cost obtained by Maslov \emph{et al.} is not linear and so is
ours. Consequently it will be completely justified to compare the quantum cost obtained by our proposed
algorithm with that obtained using  Maslov \emph{et al.}'s algorithm.

\begin{figure}[h]
\centering

\includegraphics[scale=0.7]{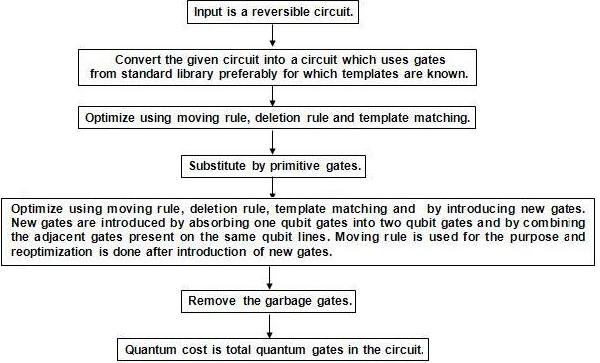}
\caption{Algorithm for minimization of quantum cost.}

\end{figure}

\section{Optimization algorithm} In this section we have proposed an algorithm that optimizes the quantum cost of classical reversible and quantum circuits. It is presented in the form of a flowchart in Fig 1. The flowchart is explained below:
\begin{figure}[h]
\centering
\includegraphics[scale=.45]{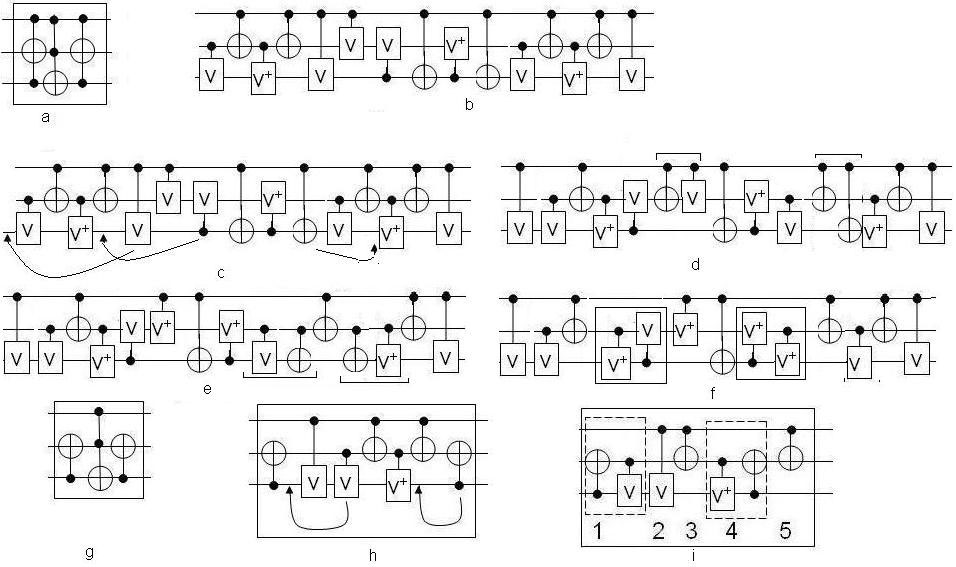}\caption{ a) A Fredkin gate is implemented using three Toffoli gates.  b) Toffoli gate is substituted by primitives, so its
direct linear cost is $5\times3=15$. c) Moving rule is applied (the movements are shown by arrows), d) and e)
template matching rule is applied.  f) New gates are introduced (dashed boxes) and the quantum cost is 11.
g) The circuit shown in Fig 2a is reduced here by template matching to one Toffoli and two Cnot gates. h) The
Toffoli gate is substituted by quantum primitives. According to Haghparast and Eshghi 's methods
\cite{Mohammadifiguesofmerit} the quantum cost is now 7. The moving rule is applied  to circuit. i)  New gates
are introduced to yield quantum cost of Fredkin gate as 5. }
\end{figure}
\begin{enumerate}
\item The input is a reversible circuit.   Here we would like to note that our goal is  to find out the minimum
number of quantum primitive gates required to implement the circuit and we are not much concerned about the
choice of gate library \cite{Maslovbenchmark, Barenco, Nielsen} in principle.  But in practice it is easier to
work using an input circuit which is constituted using the gates from a standard gate library for which a
large/complete set of templates are known. At present  there are few  set of templates, available  for classical reversible circuits
\cite{Maslov2,Maslov4}. However not much templates \cite{Maslov2} are reported for quantum circuits and it is
not difficult to generate them. Therefore, in the beginning of the algorithm we convert the input reversible
circuit into a circuit composed of gates taken from a standard gate library preferably  those gate libraries  for
which a complete/large set of templates are already known.

\item In the next step we  optimize the gate count of the reversible circuit by applying local optimization
tools  which are moving rule, deletion rule and template matching. We apply moving rule or commutation rule
\cite{Nielsen} which is simply a matrix operation to see whether the adjacent gates commute or not.  This
operation  is useful  to reduce the gate count with the help of self inverse rule and template matching
\cite{Maslov2}. If at any point of time we find that the adjacent gates are of the same type and  they form an
identity (I) then we can remove both of the identical gates. This is called self inverse or deletion rule. In
NCT gate library, all the gates are self inverse and in NCV gate library apart from the square root of Not gate
(where $v$.$v^+$ = I) all the remaining gates are self inverse. In template matching \cite{Miller} a sequence of
gates  is substituted by another sequence of gates  having lesser number of gate count and same operational
effect. Suppose we have a template: $U_{1} U_{2} U_{3} U_{4} U_{5}$=I (where $U_{i}$ is an unitary gate) and in the
optimization procedure we come across a sequence of gates $U_{2} U_{3} U_{4} $ then we can replace this sequence
of gates by $U_{1}^{-1} U_{5}^{-1}$.
\begin{figure}[h]
\centering
\includegraphics[scale=0.57]{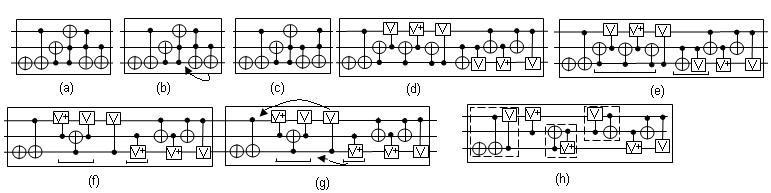}\caption{ a) Reversible circuit for function  3\_17  given in benchmark page of Maslov \emph{et al}.\cite{Maslovbenchmark}. b) Commutation rule is applied and arrow shows the movement of Cnot gate.  c) NCT circuit before substitution of primitives d) Quantum circuit of 3\_17 function obtained by substituting the Toffoli gates with primitives. e) Template matching tool is applied from \cite{Maslov2} to the circuit. f) Quantum circuit with reduced gate count. g) Modified local optimization rule is applied and two movements have been done in the circuit as indicated by the arrows. h) New gates are introduced (each box is a new gate) and quantum cost of the circuit is obtained as the total quantum gates present in the
circuit. The quantum cost of this circuit is 7. } \end{figure}

~

\item In this step we  obtain  an equivalent primitive circuit. This is done by decomposing  every n qubit gates (where n  $\geqslant$ 3) into equivalent circuit comprising of  elementary gates  ($2\times2$ or $4\times4$ quantum gates).

\item We optimize  the circuit comprising of quantum primitive gates  in the  following steps.
\begin{enumerate}
\item We apply moving rule, deletion rule and modified template matching. In modified template matching a
sequence of gates is substituted by another sequence of gates if it decreases  the overall quantum cost of the
circuit. In step 2 we have explained how a standard template matching reduces quantum cost by reducing the gate
count. In modified template matching the overall cost is reduced by simultaneous application of template
matching and introduction of new gates. Here we may substitute a sequence of gates by a larger sequence of gates
if after the substitution, the gates present at the edge of the new sequence merges with the adjacent gates on
the same qubit lines to reduce the overall quantum cost. It is explained in example 1 of this section. \item  We
club together the adjacent gate/gates of dimension $2\times2$ and  $4\times4$, $4\times4$ and  $2\times2$,
$2\times2$ and  $2\times2$, $4\times4$ and $4\times4$  to form new gates.   In the circuit there may be other
gates in the same qubit lines but not adjacent. In this step we will apply commutation rule and if the gates on
the same qubit line or lines  can be brought adjacent they will again form a new gate and reduce the cost. This
is a modified optimization where we introduce new gate and apply the commutation rule to decrease the quantum
cost of the circuit. \item Since new gates are formed in the procedure, the existing   gates in the circuit  may
belong to another gate library and it is possible that templates for that particular gate library exist, hence
we explore the  further scope of minimization of gate count by template matching and deletion rule. We may
require  generating new templates for this procedure.

\end{enumerate}
\item We  remove those gates,  which do not affect the output or in other words affect only the garbage bits.
When we substitute  Toffoli gate by quantum primitives then there appear a  lot of unnecessary quantum gates.
This situation is similar to the garbage bits which are added to make an irreversible function reversible.
Analogously these gates can be called as garbage gates.   For example, if during computation the  desired output
of the circuit is obtained from the third qubit line in Fig. 2i then  first two qubit lines at the output are
garbage bits and the last two Cnot gates are garbage gates. Another example is a reversible function like 4mod5
\cite{Maslovbenchmark} (Grovers oracle) whose output is 1 if the 4 bit input is divisible by 5. The circuit has
one desired output and rest of the output bits are garbage bits. In this case when we apply our quantum cost
minimization algorithm we find it helpful to remove those gates (garbage gates) that affect only the garbage
bits. This unique feature of quantum cost optimization algorithm is applied in the
present work to minimize the cost of 4mod5 d1 circuit%
\footnote{4mod5 d1 stands for the design number 1 given in benchmark to realize
4mod5 function from NCT gate library.%
} (see Table 1).

\item  The quantum cost of entire circuit is obtained as the total number of  quantum gates present in the circuit.

\end{enumerate}

To illustrate how this algorithm works let us consider following two examples.
\begin{enumerate}
\item Consider a Fredkin gate and convert it to NCT circuit by applying a synthesis algorithm \cite{Miller} as
shown in Fig. 2a. We will calculate its quantum cost in two parts which is without optimizing the NCT circuit
and after optimizing the NCT circuit. In the first part we substitute the Toffoli gate with its quantum
primitives as shown in Fig. 2b. In Fig.2c we have applied moving rule and indicated the movement by arrows.
There are two places as shown in Fig. 2d where modified template matching can be applied and  the resultant
circuit is shown in Fig. 2e, here we have also marked the places where we can again apply templates. We obtain a
circuit shown in Fig. 2f, we have marked in boxes the new gates and find that
 the  quantum cost is 11.  In the second part  we will optimize the NCT circuit of Fredkin gate in Fig. 2a by applying template matching and obtain an  optimized circuit as shown in Fig. 2g and further the Toffoli
gate is substituted by primitives shown in Fig. 2h. We have applied modified optimization rule (commutation is
shown by arrow) and in Fig. 2i we have shown the new gates formed. The quantum cost of the circuit is 5.
 This example clearly establishes  that it is very essential to optimize the reversible circuit before substituting it with
its quantum primitives. This aspect is not mentioned in earlier works \cite{Smolin, Maslov2, Perkowski, Hung}.
It also clearly explains the meaning of modified template matching protocol introduced in the present work.
\item The reversible NCT circuit for function 3\_17,   is shown in Fig 3a \cite{Maslovbenchmark}. This is the
input of our algorithm, in Fig 3b we have shown that the end Cnot gate will commute with adjacent Toffoli gate
(thereby reduce the quantum cost) and the movement is shown by an arrow. The resultant circuit after commutation
is shown in Fig. 3c.  We  try to optimize its gate count but we find that we cannot apply self inverse rule or
template matching. We substitute the Toffoli gate with its primitives and the resultant circuit is shown in Fig.
3d. We try to optimize the circuit,  there are two places shown in Fig. 3e where  templates can be applied and
after the application of templates we have obtained the circuit which is shown  in Fig. 3f.  Thereafter we apply
modified optimization technique  in Fig. 3g, new gates are formed  which are shown in boxes in Fig. 3h. Finally
we calculate total number of quantum gates in the circuit and find the quantum cost of the circuit.

\end{enumerate}

\subsection{Quantum cost optimized circuits }
\begin{table}
\centering
\includegraphics[scale=.75]{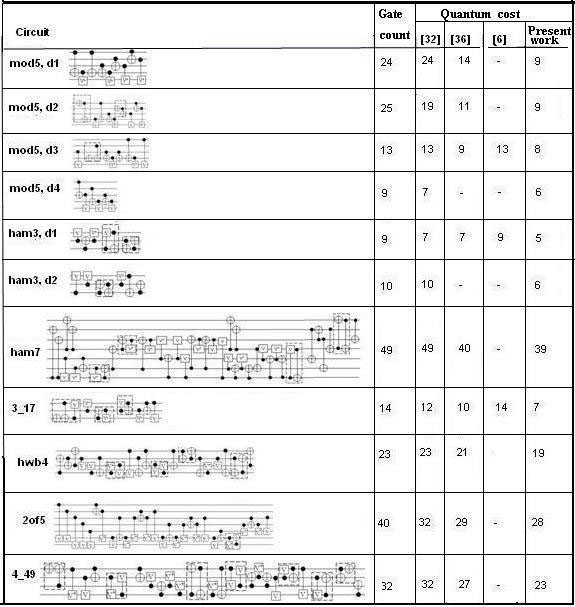}
\caption{Comparison of quantum cost using our algorithm with the existing works of Maslov \emph{et al.}
\cite{Maslovbenchmark}, Maslov \emph{et al.} \cite{Maslov2} and Gupta \emph{et al.} \cite{Gupta}. }
\end{table}
\begin{table}
\centering
\includegraphics[scale=0.7]{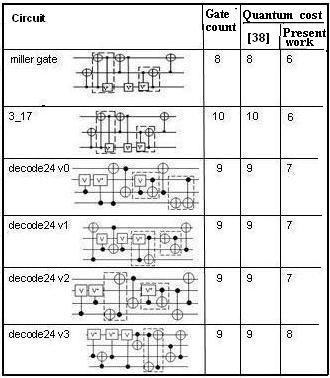}
\caption{Comparison of quantum cost using our algorithm with the existing works of Revlib \cite{Revlib}.}
\end{table}
\begin{table}
\centering
\includegraphics[scale=0.55]{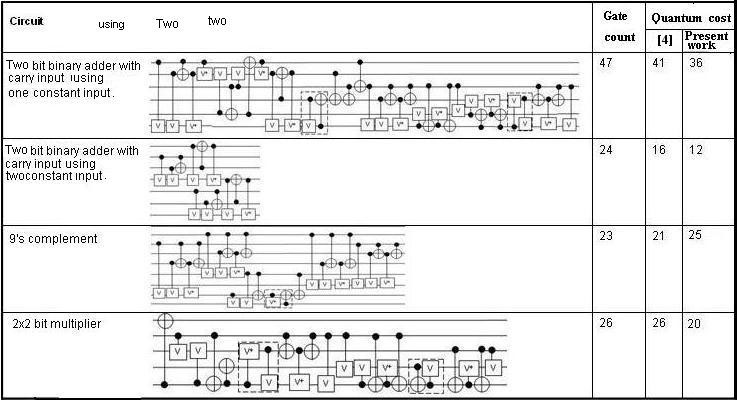}\caption{Comparison of quantum cost using our algorithm with the existing works
of Mohammadi \cite{Mohammadifiguesofmerit}.}
\end{table}

\begin{table}
\centering
\includegraphics[scale=0.38]{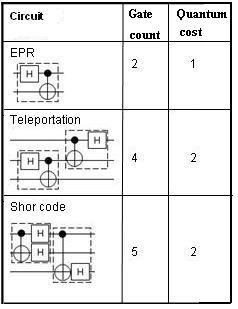}\caption{Quantum cost of important quantum circuits.}
\end{table}

We have already mentioned that most of the existing results related to quantum cost are available in benchmark
page of Maslov \emph{et al.} \cite{Maslovbenchmark} and in Revlib \cite{Revlib}. In addition to these two databases Mohammadi and
 Eshghi \cite{Mohammadifiguesofmerit}, Gupta \emph{et al.} \cite{Gupta} and Maslov \emph{et al.}\cite{Maslov2} have independently reported the quantum cost of different reversible circuits. We have compared the quantum costs reported in these works with the quantum costs of the same functions obtained using the present algorithm. The results of comparison are shown in Table 1 - Table 3. To be precise in Table 1 we have compared the quantum costs of the following functions: i) mod5
function which is divisibility checker, ii) ham3 which is the size 3 hamming optimal coding function, iii) ham7
which is size 7 hamming optimal coding function, iv) hwb4 which is the hidden weighted bit function \cite{HWB}
with parameter N=4, v) 3\_17 which is the worst case scenario 3 variable function \cite{Miller} having function
specification \{7, 1, 4, 3, 0, 2, 6, 5\} and vi) 4\_49 which is the worst case scenario 4 variable function
\cite{Miller} having function specification \{15, 1, 12 ,3 ,5 ,6 ,8 ,7 ,0 ,1 0, 13 ,9 ,2 ,4 ,1 4, 11\}. Here we
would like to note that in this paper we have  mentioned the circuits described in \cite{Maslovbenchmark} as
benchmark circuits. To provide specific examples and to establish the superiority of our algorithm we have
applied our algorithm to those benchmark circuits.   Further, the benchmark circuits reported in
\cite{Maslovbenchmark} to realize a particular function is not unique and consequently different designs for the
same purpose are marked with different indices, for example d1 denotes design 1, d2 denotes design 2 etc. Here
we have followed the same convention as it is used in \cite{Maslovbenchmark}. Gupta \emph{et al.} \cite{Gupta}
have synthesized few reversible circuits for realization of above mentioned functions
  in the form of a  network of Toffoli gates and  have also
reported their quantum costs. Further in \cite{Maslov2} improved quantum costs are reported for various circuits
reported earlier  \cite{Maslovbenchmark}. We have compared the quantum costs reported in these works in Table 1.

~

~
 In Table 2 we have reported quantum cost of circuits from  Revlib \cite{Revlib}. To be precise, we have
compared quantum cost of the following functions: i) miller gate, ii) 3\_17 which is the worst case scenario 3
variable function \cite{Miller} and iii) different designs of  decode 24 function which  is 2 to 4 binary
decoder. Table 3 compares quantum costs of some circuits that has been reported in
\cite{Mohammadifiguesofmerit}. For example: i)
  two bit binary adder with carry input using one constant input (see Fig. 3a of \cite{Mohammadifiguesofmerit}), ii)
 two bit binary adder with carry input using two constant input (see Fig. 3b of \cite{Mohammadifiguesofmerit}), iii) 9's complement circuit without constant inputs (see Fig. 4a of \cite{Mohammadifiguesofmerit}) and iv)  $2\times2$ bit multiplier (see Fig. 15 of \cite{Mohammadifiguesofmerit}).
 The algorithm may be applied to other benchmark circuits too but to do so either one has to develop templates for the corresponding gate library or convert the circuit into  other gate library for which templates has been provided in literature example NCT gate library. In Table 4, we have  calculated quantum cost of some pure quantum circuits like EPR, quantum teleportation and
shor code. Since quantum cost of these circuits have not been reported earlier, therefore its comparison could
not be done. The quantum cost optimized circuits are shown in the first column of Table 1 - Table 4. The gates
shown in the dotted box should form a new gate and it would be counted as a single gate in the calculation of
quantum cost. ~

\section{Conclusions}

We have proposed an algorithm for minimization of quantum cost. We have applied our algorithm to different
circuits from various sources \cite{Mohammadifiguesofmerit, Gupta, Maslovbenchmark, Maslov2, Revlib} and
compared our results.  The outcome of the comparison (see Table 1 - Table 3) clearly shows that the proposed
algorithm produces best  result. In Table 4 we have reported  quantum cost of different quantum circuits (for
example, quantum teleportation, EPR circuit etc.).  Through these examples it is clearly established that the
proposed algorithm is useful in reduction of quantum cost. Thus the present algorithm provides a window for
reduction of quantum cost of other circuits in future.

~

\end{document}